\documentclass[aps,prb,twocolumn,showpacs]{revtex4} 
\usepackage{amsmath}
\usepackage{graphicx}
\usepackage{bm}

\begin{document}

\title{Two-dimensional arrays of superconducting strips as dc magnetic metamaterials}

\author{Yasunori Mawatari$^1$, Carles Navau$^2$, and Alvaro Sanchez$^2$}

\affiliation{$^1$%
National Institute of Advanced Industrial Science and Technology (AIST), 
Tsukuba, Ibaraki 305-8568, Japan}
\affiliation{$^2$%
Grup d'Electromagnetisme, Departament de F\'{\i}sica, 
Universitat Aut\`onoma de Barcelona, 08193 Bellaterra,
Barcelona, Catalonia, Spain}

\date{December 1, 2011}
\pacs{74.25.Ha, 74.78.Fk, 81.05.Xj}%

\begin{abstract} 
We theoretically investigate the magnetic response of two-dimensional arrays of superconducting strips, which are regarded as essential structures of dc magnetic metamaterials. 
We analytically obtain local distributions of the magnetic field for the ideal complete shielding state (i.e., $\Lambda/w\to 0$, where $2w$ is the strip width, $\Lambda=\lambda^2/d$ is the Pearl length, $\lambda$ is the London penetration depth, and $d$ is the strip thickness), and derive effective permeability by averaging the local field distributions. 
We also perform numerical calculations for a realistic case, taking finite $\Lambda/w>0$ into account. 
We investigate two types of strip arrays: a rectangular array and a hexagonal array. 
The resulting effective permeability has large anisotropy that depends on the dimensions and arrangement of the superconducting strips, and the hexagonal array is found to be more advantageous for obtaining large anisotropy than the rectangular array. 
\end{abstract}

\maketitle

\section{Introduction}
The London model describes the macroscopic electromagnetic behavior of a superconductor when fields and current are weak and vary on a length scale $\lambda$ (the London penetration depth).~\cite{degennes} In many cases, the superconducting sample has large dimensions such that the currents flow only along the surface of the sample, which is completely shielded from the external magnetic field. This complete shielding state corresponds to the limit $\lambda\rightarrow 0$. Another macroscopic model for describing superconductors with strong vortex pinning is the critical state model.~\cite{bean} This model assumes that the current density induced inside a superconductor cannot exceed the critical current density $J_c$. The complete shielding state is also observed at the high $J_c$ (or low applied field) limit in the critical state.~\cite{bean,brandteurophysics,clemsanchez} 

In addition to the trivial cases of infinite cylinders or slabs in a longitudinal applied field, distributions of surface current density in flat superconducting samples under uniform transverse magnetic fields can be analytically determined for the complete shielding state in several geometries, such as thin disks or strips,~\cite{landau} coplanar pairs of thin strips,~\cite{brojeny02sust15} periodically arranged thin strips,~\cite{mawatari96prb54} and thick infinite tapes~\cite{chen02jap91}. In Ref.~\onlinecite{mawatari96prb54}, the exact analytical expressions were presented for the sheet current density and magnetic field profiles in an infinite stack and an infinite coplanar array of thin superconducting strips in the complete shielding state. Numerically, the sheet currents can also be calculated for thin plates of arbitrary shape in a perpendicular applied field.~\cite{brandt95prl74,brojeny03prb68,prigozhin96jcomp129,navau08jap103} 

Within the London model taking finite $\lambda>0$ into account, complete shielding is not obtained because the magnetic field penetrates the superconductor. 
In a superconducting strip of thickness $d$ and width $2w$ (where $d\ll 2w$), the relevant length scale for magnetic field penetration is the Pearl length $\Lambda=\lambda^2/d$.~\cite{Pearl64,Larkin72} 
Large current flows near the edges of a wide strip with $\Lambda/w\ll 1$,~\cite{Brandt93b,Zeldov94,Dorsey95} whereas the distribution of the sheet current for a narrow strip $\Lambda/w\gg 1$ is simply proportional to the distance from the center.~\cite{Dorsey95} 
There are, however, no exact analytical expressions for strips of arbitrary width in perpendicular fields. 
Numerical calculations with $\lambda>0$ can account for the sheet current density in some thin geometries under perpendicular applied fields.~\cite{chen08sust21,clem05prb72,brandt04prb69,brandt01prb64,plourde02prb64} 

The above-mentioned results have been widely used as theoretical frameworks for investigating superconductors, ranging from studies of flux penetration in mesoscopic superconductors to the macroscopic characterization of bulk superconductors. In the present paper, we extend the previous results by considering two-dimensional arrays of superconducting thin strips exposed to perpendicular fields, and we offer a theoretical framework for a new class of superconducting systems, dc magnetic metamaterials.

Metamaterials have recently attracted considerable attention, because they can provide effective (or macroscopic) electromagnetic properties not found in nature, allowing a new set of applications for controlling electromagnetic field.~\cite{pendry96prl76,pendry06science312} 
These effective electromagnetic properties come from the internal structure of the metamaterials. Typically, metamaterials are composed of an array of structure whose sizes and separation from each other are smaller than the wavelength of the involved electromagnetic field. Tailoring such internal structure is the key factor in obtaining the desired effective properties. A crowning achievement would be electromagnetic field cloaking. One can consider, for example, a spherical metamaterial with a concentric spherical hole exposed to an incident electromagnetic wave; inside the metamaterial sphere the total electromagnetic field is zero, whereas outside the metamaterial sphere the electromagnetic field is undisturbed. 
Effectively, one can find the values of the permittivity and permeability tensors that ensure the cloaking of electromagnetic fields from transformation optics,~\cite{pendry06science312} conformal mapping,~\cite{leonhardt03ie3jstqe9} or by considering boundary conditions.~\cite{yaghjian08njp10} However, the exact cloaking of a broadband electromagnetic field may involve unphysical situations, such as a diverging phase-velocity of light or violations of causality-energy conditions.~\cite{yaghjian08njp10} Moreover, extreme values of the permittivity and permeability are needed and, in most cases, the use of resonant structures results in large losses.~\cite{schurig06science314} 

The control of a dc magnetic field, either for cloaking or for other possible applications, represents the zero frequency limit of the general electromagnetic field. In this dc field, the electric and magnetic fields are decoupled and only the magnetic permeability of the material is relevant.~\cite{wood07jpcm19} However, the control of the magnetic field requires metamaterials that should be, in general, anisotropic and inhomogeneous (i.e., the permeability tensor of the material must be a particular function of the position), depending on the desired type of control. 

Arrays of thin superconductors such as dc superconducting metamaterials~\cite{wood07jpcm19} can partially provide such characteristics, since they are intrinsically anisotropic. Indeed, maintaining parallel permeability $\mu_\parallel=\mu_0$ (where $\mu_0$ is the vacuum permeability), the perpendicular effective permeability $\mu_\perp$ can be tuned from $0$ to $\mu_0$ by changing the geometric parameters of the array.~\cite{navau09apl94} The first experimental demonstration of such dc superconducting metamaterials providing $\mu_\perp/\mu_0<1$ was presented in Ref.~\onlinecite{magnus08natmat7}, using arrays of thin Pb films. 
The physics and applications of superconducting metamaterials (including dc superconducting metamaterials) were recently reviewed in Ref.~\onlinecite{anlage11jopt13}. 
Superconductor-metamaterial hybrids have also been proposed as antimagnets.~\cite{Sanchez11} 

In the present work we present analytical expressions for the magnetic field distribution, as well as for the effective permeability of metamaterials consisting of infinite regular two-dimensional (2D) arrays of superconducting strips in the complete shielding state with $\Lambda/w\to 0$. Such analytical results are useful because they provide general trends in the relation between the effective magnetic properties of a metamaterial and the geometry of its constituents. 
Moreover, we numerically calculate how these effective properties change when a nonzero $\Lambda/w$ is considered in the superconducting strips. 

This paper is organized as follows. 
After an introduction to 2D arrays of superconducting strips in Sec.~\ref{sec:2D-array}, we theoretically investigate two types of 2D arrays of superconducting strips: a rectangular array in Sec.~\ref{sec:rectangular-array}, and a hexagonal array in Sec.~\ref{sec:hexagonal-array}. 
In Secs.~\ref{sec:rectangular-array} and \ref{sec:hexagonal-array}, we present an analytical investigation of the field distributions and effective permeability for the case of complete shielding with $\Lambda/w\to 0$, and also a numerical investigation on the effective permeability for the case of finite penetration depth with $\Lambda/w>0$. 
In Sec.~\ref{sec:discussion} our theoretical results are discussed and a brief summary is given.

\section{Two-dimensional arrays of superconducting strips
\label{sec:2D-array}}
We now describe 2D arrays of superconducting strips as basic components of magnetic metamaterials. 
The width of the superconducting strips $2w$ is much larger than their thickness $d$, and their length is infinite along the $z$ axis. 
The wide surfaces of superconducting strips are parallel to the $xz$ plane. 
We consider the thin-strip limit case, $\epsilon=d/2\to 0$, and hereafter we regard $\epsilon\to 0$ to be an infinitesimal. 

The relation between the local (or microscopic) magnetic field $\bm H$ and the local magnetic induction $\bm B$ is given by ${\bm B}=\mu_0{\bm H}$. 
The macroscopic fields ${\langle\bm H \rangle}$ and ${\langle\bm B \rangle}$ are obtained by respectively averaging $\bm H$ and $\bm B$ in the unit cell of the strip array. 
As shown in Refs.~\onlinecite{Pendry99mtt47} and \onlinecite{Smith06jos23}, to maintain consistency with the Maxwell equations ${\langle\bm H \rangle}$ is calculated as the averaged line integral of $\bm H$, whereas ${\langle\bm B \rangle}$ is calculated as the averaged surface integral of $\bm B$ in the unit cell.
Because of the different definitions of averaging procedure to obtain macroscopic fields, we generally have ${\langle\bm B \rangle}\neq\mu_0{\langle\bm H \rangle}$, even though ${\bm B}=\mu_0{\bm H}$ holds. 

The macroscopic magnetic response of the 2D arrays of superconducting strips is characterized by the relation between the macroscopic magnetic field ${\langle\bm H \rangle}= {\langle H_x \rangle}\hat{\bm x} +{\langle H_y \rangle}\hat{\bm y}$ and the macroscopic magnetic induction ${\langle\bm B \rangle}= \langle B_x \rangle\hat{\bm x} +\langle B_y \rangle\hat{\bm y}$, as $\langle B_{\alpha} \rangle= \mu_{\alpha\beta} \langle H_{\beta} \rangle$, where the components of the permeability tensor are $\mu_{xx}=\mu_{\parallel}$, $\mu_{yy}=\mu_{\perp}$, and $\mu_{\alpha\beta}=0$ for $\alpha\neq\beta$. 
In other words, 
\begin{equation}
	\langle B_x \rangle=\mu_{\parallel}\langle H_x \rangle 
	\quad\mbox{and}\quad
	\langle B_y \rangle=\mu_{\perp}\langle H_y \rangle . 
\label{eq:macro_B-H}
\end{equation}

When a 2D array of superconducting strips is exposed to a parallel magnetic field (i.e, one along the $x$ axis), the field is not disturbed by thin strips of $\epsilon\to 0$. 
The effective permeability for the parallel field is, therefore, equal to the vacuum permeability $\mu_{\parallel}=\mu_0$ for the thin strip limit. 

On the other hand, when a 2D array of superconducting strips is exposed to a perpendicular magnetic field (one along the $y$ axis), the strips strongly disturb the field. 
Because of the magnetic shielding in superconducting strips, the effective permeability for the perpendicular field is smaller than the vacuum permeability, $\mu_{\perp}<\mu_0$, depending on the geometrical parameters for the 2D array, $2a$, $2b$, and $2w$. 

Hereafter we study field distributions and effective perpendicular permeability $\mu_{\perp}$ for the case where 2D arrays of superconducting strips are exposed to perpendicular magnetic fields.

\section{Rectangular arrays of superconducting strips
\label{sec:rectangular-array}}

\begin{figure}[t]
	\includegraphics{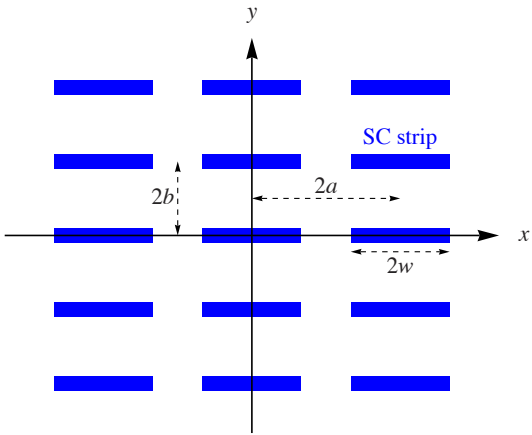}
\caption{(Color online) %
Cross section of a two-dimensional array of superconducting strips in the $xy$ plane. 
In the $n$th layer at $y=2nb$, the $m$th strip is situated at $|x-2ma|<w$, where $m=0,\,\pm 1,\,\pm 2,\ldots\pm\infty$ and $n=0,\,\pm 1,\,\pm 2,\ldots\pm\infty$. 
}
\label{fig:rectangular-array}
\end{figure}

Figure~\ref{fig:rectangular-array} shows a schematic of the rectangular array of superconducting strips investigated in this section. 
Superconducting strips are regularly arranged with a unit cell of $2a\times 2b$ in the $xy$ plane; the lattice constant along the $x$ axis is $2a$, and that along the $y$ axis is $2b$, where $2a>2w\gg d$ and $2b\gg d$. 

\subsection{Complete shielding state, $\Lambda/w\to 0$}
Here we consider field distributions and perpendicular permeability of a rectangular array of superconducting strips for the complete shielding state (i.e., $\Lambda/w\to 0$). 
Such complete shielding can also be seen for a weak-field or large-$J_c$ limit (i.e., $H_a/J_cd\to 0$) in the critical state model, where $H_a$ is the applied magnetic field and $J_c$ is the critical current density.

\subsubsection{Field distribution}
The two-dimensional local magnetic field $\bm{H}=H_x(x,y)\hat{\bm{x}} +H_y(x,y)\hat{\bm{y}}$ is analytically investigated via the complex field ${\cal H}(\zeta)=H_y(x,y) +iH_x(x,y)$ as an analytic function of $\zeta=x+iy$.~\cite{landau,Beth66jap37}

We use the conformal mapping from the $\zeta=x+iy$ plane to $\eta=u+iv$ plane,~\cite{Kober} 
\begin{equation}
	\eta=\eta_{\rm r}(\zeta)\equiv \mbox{sn}(\zeta/c_{\rm r},k_{\rm r}) , 
\label{eq:zeta-eta}
\end{equation}
where $\mbox{sn}(u,k)$ is the sine amplitude (the Jacobi sn function).~\cite{Gradshtein-Ryzhik} 
[See Eq.~\eqref{eq:sn}.] 
The modulus $k_{\rm r}$ is the function of $b/a$, and is obtained by solving 
\begin{equation}
	\frac{b}{a}= \frac{{\bf K}(k_{\rm r}')}{{\bf K}(k_{\rm r})} , 
\label{eq:b/a-K}
\end{equation}
where ${\bf K}(k)$ is the complete elliptic integral of the first kind~\cite{Gradshtein-Ryzhik} [see Eq.~\eqref{eq:K_comp-ellip-int}] and $k_{\rm r}'=\sqrt{1-k_{\rm r}^2}$. 
The $c_{\rm r}$ in Eq.~\eqref{eq:zeta-eta} is then given by 
\begin{equation}
	c_{\rm r} =a/{\bf K}(k_{\rm r}) =b/{\bf K}(k_{\rm r}') . 
\label{eq:c=a/K=b/K'}
\end{equation}
The upper half of the unit cell in the $\zeta$ plane, $|x|<a$ and $0<y<b$, is mapped onto the upper half plane in the $\eta$ plane, $0<v<\infty$.~\cite{Kober} 
From Eqs.~\eqref{eq:sn_u=0}--\eqref{eq:sn_u=iK'}, we see that the points $\zeta=0$, $a$, $a+ib$, and $ib$ in the $\zeta$ plane are mapped onto $\eta=0$, $1$, $1/k_{\rm r}$, and $\infty$ in the $\eta$ plane, respectively. 

Using the conformal mapping technique, we obtain the complex field for the rectangular array of superconducting strips in the complete shielding state, \begin{equation}
	{\cal H}(\zeta)= H_0\frac{\eta_{\rm r}(\zeta)}{%
		\sqrt{\eta_{\rm r}(\zeta)^2- \eta_{\rm r}(w)^2}} . 
\label{eq:H(z)}
\end{equation}
Note that the coefficient $H_0=H_y(0,b)$ in Eq.~\eqref{eq:H(z)} is the magnetic field at $(x,y)=(0,b)$. 
For $b/a\to 0$, we have $k_{\rm r}\to 1$ and $\eta_{\rm r}(\zeta)\to\tanh(\pi\zeta/2b)$ from Eq.~\eqref{eq:sn_k=1}, which corresponds to the case of an infinite stack of superconducting strips.~\cite{mawatari96prb54} 
For $b/a\to \infty$, on the other hand, we have $k_{\rm r}\to 0$ and $\eta_{\rm r}(\zeta)\to\sin(\pi\zeta/2a)$ from Eq.~\eqref{eq:sn_k=0}, which corresponds to the case of an infinite array of coplanar superconducting strips.~\cite{mawatari96prb54} 

The perpendicular field component $H_y(x,0)={\rm Re}\,{\cal H}(x)$ and the sheet current $K_z(x)=H_x(x,-\epsilon)-H_x(x,\epsilon)={\rm Im}\,[{\cal H}(x-i\epsilon)-{\cal H}(x+i\epsilon)]$ 
in the unit cell $-a<x<a$ are obtained from Eq.~\eqref{eq:H(z)}:
\begin{eqnarray}
	H_y(x,0)&=& 
		\begin{cases} 
		\ 0 & \mbox{for } |x|<w , \\[1em]
		\displaystyle \frac{H_0\,\eta_{\rm r}(x)}{%
		\sqrt{\eta_{\rm r}(x)^2-\eta_{\rm r}(w)^2}} 
		& \mbox{for } w<|x|<a , 
	\end{cases}
\label{eq:hy(x,0)}\\
	K_z(x)&=& 
		\begin{cases} 
		\displaystyle \frac{2H_0\,\eta_{\rm r}(x)}{%
		\sqrt{\eta_{\rm r}(w)^2-\eta_{\rm r}(x)^2}} 
		& \mbox{for } |x|<w , \\[1.5em]
		\ 0 & \mbox{for } w<|x|<a , 
	\end{cases}
\label{eq:Kz(x)}
\end{eqnarray}
where $\eta_{\rm r}(x)=\mbox{sn}(x/c_{\rm r},k_{\rm r})$. 

\begin{figure*}[bt]
	\includegraphics{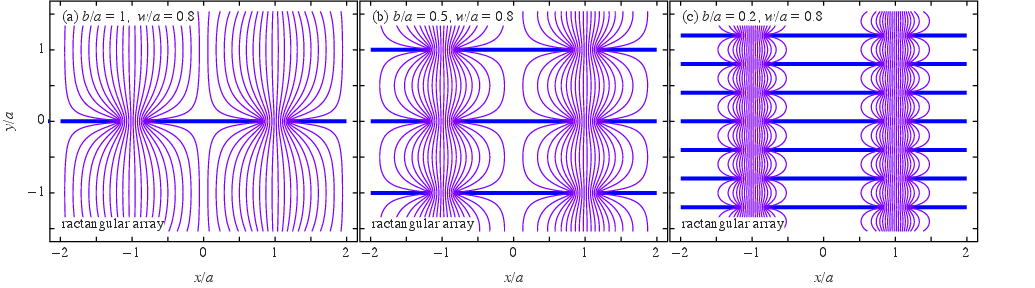}
\caption{(Color online) %
Magnetic field lines in a rectangular array of superconducting strips for $w/a=0.8$: (a) $b/a=1$, (b) $b/a=0.5$, and (c) $b/a=0.2$. 
Thick horizontal lines correspond to the cross sections of superconducting strips. 
}
\label{fig:field-lines}
\end{figure*}

The complex potential defined by ${\cal G}(\zeta)= \int_{ib}^{\zeta}{\cal H}(\zeta')d\zeta'$ is calculated from Eq.~\eqref{eq:H(z)}, and is given by 
\begin{eqnarray}
	{\cal G}(\zeta) &=&
	\frac{c_{\rm r}H_0}{k_{\rm r}} \int_{\eta_{\rm r}(\zeta)}^{\infty} 
		\frac{udu}{\sqrt{(u^2-1)(u^2-k_{\rm r}^{-2})(u^2-\gamma_{\rm r}^2)}} 
\nonumber\\
	&=& \frac{c_{\rm r}H_0}{\sqrt{1-k_{\rm r}^2\gamma_{\rm r}^2}} 
		F\left(\arcsin\sqrt{\frac{k_{\rm r}^{-2} 
		-\gamma_{\rm r}^2}{\eta_{\rm r}(\zeta)^2-\gamma_{\rm r}^2}}, 
		\kappa_{\rm r}\right) , 
	\quad
\label{eq:G(z)}
\end{eqnarray}
where $F(\varphi,k)$ is the elliptic integral of the first kind.~\cite{Gradshtein-Ryzhik} 
[See Eq.~\eqref{eq:F_elliptic-int}.] 
The parameters $\gamma_{\rm r}$ and $\kappa_{\rm r}$ in Eq.~\eqref{eq:G(z)} are defined by 
\begin{eqnarray}
	\gamma_{\rm r} &=& \eta_{\rm r}(w) =\mbox{sn}(w/c_{\rm r},k_{\rm r}) , 
\label{eq:gamma}\\
	\kappa_{\rm r} &=& 
	\sqrt{\frac{1-\gamma_{\rm r}^2}{k_{\rm r}^{-2}-\gamma_{\rm r}^2}} 
	= k_{\rm r} \mbox{cd}\,(w/c_{\rm r},k_{\rm r}) , 
\label{eq:k2}
\end{eqnarray}
where $\mbox{cd}(u,k)=\mbox{cn}(u,k)/\mbox{dn}(u,k)$ is the Jacobi cd function (see Appendix~\ref{sec:elliptic-function}). 
Figure~\ref{fig:field-lines} shows the magnetic field lines corresponding to the contour lines of ${\rm Re}\,{\cal G}(x+iy)$ obtained from Eq.~\eqref{eq:G(z)}. 
The magnetic field concentrates near the gaps between the edges of superconducting strips, and this field concentration is severe for small stack spacings $b/a\ll 1$, as in Fig.~\ref{fig:field-lines}(c).

\subsubsection{Macroscopic fields}
The local magnetic field $\bm{H}=H_x(x,y)\hat{\bm{x}} +H_y(x,y)\hat{\bm{y}}$ is obtained from Eq.~\eqref{eq:H(z)}, and the local magnetic induction is given by $\bm{B}=\mu_0\bm{H}$. 
Here we consider the macroscopic magnetic field $\langle\bm{H}\rangle$ and the macroscopic magnetic induction $\langle\bm{B}\rangle$. 
Because we are interested in the case where the applied magnetic field is parallel to the $y$ axis, only $y$ components of the averaged fields $\langle H_y\rangle$ and the macroscopic magnetic induction $\langle B_y\rangle$ are relevant. 
The effective permeability is then $\mu_{\perp} =\langle B_y\rangle/\langle H_y\rangle$. 

The macroscopic magnetic induction is calculated from the averaged surface integral of the local magnetic induction in the unit cell.~\cite{Pendry99mtt47,Smith06jos23}
We therefore define $\langle B_y\rangle$ as the averaged surface integral of $B_y=\mu_0H_y$ in the region of $-a<x<a$ and $0<z<L_z$ at $y=b$: 
\begin{eqnarray}
	\langle B_y\rangle 
	&\equiv&\frac{1}{2aL_z} \int_{-a}^{+a}dx \int_{0}^{L_z}dz B_y(x,b) 
\nonumber\\
	&=& \frac{\mu_0}{2a}\int_{-a}^{+a} H_y(x,b)dx . 
\label{eq:<By>_def}
\end{eqnarray}
The volume integral of $\nabla\cdot\bm{B}=0$ in $V(y_0)$, where $V(y_0)$ denotes the region of $-a<x<a$, $y_0<y<b$, and $0<z<L_z$, reduces to 
\begin{eqnarray}
	0 &=& \int_{V(y_0)} \nabla\cdot\bm{B}\, dV 
	= \int_{S(y_0)} \bm{B}\cdot d\bm{S} 
\nonumber\\
	&=& \int_{-a}^{+a}dx \int_0^{Lz}dz 
		\bigl[ B_y(x,b)-B_y(x,y_0) \bigr] , 
\label{eq:By_V-int}
\end{eqnarray}
where the surface integrals of $B_x$ at $x=\pm a$ make no contribution because $B_x(\pm a,y)=0$. 
Substituting Eq.~\eqref{eq:By_V-int} into Eq.~\eqref{eq:<By>_def} shows that 
\begin{equation}
	\langle B_y\rangle=\frac{\mu_0}{2a}\int_{-a}^a H_y(x,y_0)dx 
\label{eq:macro-By_y0}
\end{equation}
holds for any $y_0$; that is, the right-hand side of Eq.~\eqref{eq:macro-By_y0} is independent of $y_0$. 

The macroscopic magnetic field is calculated from the averaged line integral of the local magnetic field in the unit cell.~\cite{Pendry99mtt47,Smith06jos23} 
We therefore define $\langle H_y\rangle$ as the averaged line integral of $H_y$ in the region of $-b<y<b$ at $x=a$:  
\begin{equation}
	\langle H_y\rangle \equiv \frac{1}{2b} \int_{-b}^{+b} H_y(a,y)dy . 
\label{eq:<Hy>_def}
\end{equation}
The surface integral of $\nabla\times\bm{H}=\bm{J}$ in $S_z(x_0)$, where $S_z(x_0)$ denotes the region of $x_0<x<a$ and $-b<y<b$ at $z=0$, reduces to 
\begin{eqnarray}
	\int_{x_0}^a K_z(x)dx 
	&=& \int_{S_z(x_0)}(\nabla\times\bm{H})\cdot d\bm{S} 
\nonumber\\
	&=& \int_{-b}^{+b} \bigl[ H_y(a,y)-H_y(x_0,y) \bigr]dy , 
\label{eq:Hy_S-int}
\end{eqnarray}
where the line integrals of $H_x$ at $y=\pm b$ make no contribution because $H_x(x,\pm b)=0$. 
Equation~\eqref{eq:Hy_S-int} leads to 
\begin{eqnarray}
	\langle H_y\rangle 
	&=& \frac{1}{4ab} \int_{-a}^{+a}dx \int_{-b}^{+b}dy\, H_y(a,y) 
\nonumber\\
	&=& \frac{1}{4ab} \int_{-a}^{+a}dx 
		\left[ \int_{-b}^{+b} H_y(x,y)dy +\int_x^a K_z(x')dx' \right] . 
\nonumber\\
\label{eq:macro-Hy_x}
\end{eqnarray}
Substituting Eq.~\eqref{eq:macro-By_y0} into Eq.~\eqref{eq:macro-Hy_x} yields 
\begin{eqnarray}
	\langle H_y\rangle 
	&=& \frac{1}{2b}\int_{-b}^{+b} \frac{1}{\mu_0}\langle B_y\rangle dy 
		+\frac{1}{4ab} \int_{-a}^{+a}dx \int_x^a dx' K_z(x') 
\nonumber\\
	&=& \frac{1}{\mu_0}\langle B_y\rangle 
		+\frac{1}{4ab} \int_{-a}^{+a} (x'+a)K_z(x')dx' . 
\label{eq:macro-Hy-By}
\end{eqnarray}
Equation~\eqref{eq:macro-Hy-By} reduces to 
\begin{equation}
	\langle B_y\rangle/\mu_0 
	= \langle H_y\rangle + \langle M_y\rangle , 
\label{eq:By-Hy=My}
\end{equation}
where we used $\int_{-a}^{+a}K_z(x)dx=0$, and $\langle M_y\rangle$ is the magnetization defined by 
\begin{equation}
	\langle M_y\rangle \equiv 
	{}-\frac{1}{4ab}\int_{-a}^{+a} xK_z(x)dx . 
\label{eq:My}
\end{equation}
Thus, we have confirmed that the definitions of the macroscopic fields given by Eqs.~\eqref{eq:<By>_def} and \eqref{eq:<Hy>_def} are consistent with the relation between the macroscopic fields given by Eq.~\eqref{eq:By-Hy=My}.

\subsubsection{Perpendicular permeability}
The macroscopic magnetic induction is calculated by substituting Eq.~\eqref{eq:H(z)} into Eq.~\eqref{eq:<By>_def}:
\begin{eqnarray}
	\frac{1}{\mu_0}\langle B_y \rangle 
	&=& \frac{1}{a}\int_0^a H_y(x,b) dx 
	= \frac{1}{a}\int_{ib}^{a+ib} {\cal H}(\zeta) d\zeta 
\nonumber\\
	&=& \frac{c_{\rm r}H_0}{k_{\rm r} a}\int_{1/k_{\rm r}}^{\infty} 
		\frac{udu}{\sqrt{(u^2-k_{\rm r}^{-2})(u^2-1)(u^2-\gamma_{\rm r}^2)}} 
\nonumber\\
	&=& \frac{c_{\rm r}H_0}{a\sqrt{1-k_{\rm r}^2\gamma_{\rm r}^2}} {\bf K}(\kappa_{\rm r}) , 
\label{eq:<By>_cal}
\end{eqnarray}
where $\gamma_{\rm r}$ and $\kappa_{\rm r}$ are defined by Eqs.~\eqref{eq:gamma} and \eqref{eq:k2}, respectively. 
The macroscopic magnetic field is calculated by substituting Eq.~\eqref{eq:H(z)} into Eq.~\eqref{eq:<Hy>_def}:
\begin{eqnarray}
	{\langle H_y \rangle} 
	&=& \frac{1}{b}\int_0^b H_y(a,y) dy 
	= -\frac{i}{b}\int_{a}^{a+ib} {\cal H}(\zeta) d\zeta 
\nonumber\\
	&=& \frac{c_{\rm r}H_0}{k_{\rm r} b}\int_{1}^{1/k_{\rm r}} 
		\frac{udu}{\sqrt{(k_{\rm r}^{-2}-u^2)(u^2-1)(u^2-\gamma_{\rm r}^2)}} 
\nonumber\\
	&=& \frac{c_{\rm r}H_0}{b\sqrt{1-k_{\rm r}^2\gamma_{\rm r}^2}} 
		{\bf K}(\kappa_{\rm r}') , 
\label{eq:<Hy>_cal}
\end{eqnarray}
where $\kappa_{\rm r}'=\sqrt{1-\kappa_{\rm r}^2}=\sqrt{(1-k_{\rm r}^2)/(1-k_{\rm r}^2\gamma_{\rm r}^2)}$. 
The effective permeability $\mu_{\rm\perp r}=\langle B_y \rangle/{\langle H_y \rangle}$ is therefore obtained as 
\begin{equation}
	\frac{\mu_{\rm\perp r}}{\mu_0} 
	= \frac{b}{a} 
	\frac{{\bf K}(\kappa_{\rm r})}{{\bf K}(\kappa_{\rm r}')} . 
\label{eq:<mu_y>}
\end{equation}

When the width of superconducting strips is small ($w/c_{\rm r}\ll 1$), we have $\kappa_{\rm r}\simeq k_{\rm r}[1-(1-k_{\rm r}^2)w^2/2c_{\rm r}^2]$, and Eq.~\eqref{eq:<mu_y>} reduces to 
\begin{equation}
	\frac{\mu_{\rm\perp r}}{\mu_0}
	\simeq 1-\frac{\pi w^2}{4ab} . 
\label{eq:<mu_y>_small-width}
\end{equation}

When the gaps between the edges of the superconducting strips are small ($1-w/a\ll 1$), we have $\kappa_{\rm r}\simeq k_{\rm r}{\bf K}(k_{\rm r})(1-w/a)$ and ${\bf K}(\kappa_{\rm r}')/{\bf K}(\kappa_{\rm r})\simeq (2/\pi)\ln(4/\kappa_{\rm r})$, and Eq.~\eqref{eq:<mu_y>} reduces to 
\begin{equation}
	\frac{\mu_{\rm\perp r}}{\mu_0}
	\simeq \frac{\pi b}{2 a}\left[\ln\left(%
		\frac{4}{k_{\rm r}{\bf K}(k_{\rm r})(1-w/a)} \right)\right]^{-1} . 
\label{eq:<mu_y>_small-gap}
\end{equation}
The inverse of $\mu_{\rm\perp r}/\mu_0$ logarithmically diverges when $1-w/a\to 0$. 
This sharp change in $\mu_{\rm\perp r}$ for small gaps becomes gradual when we take finite $\Lambda/w$ into account, as shown in the numerical results in the next subsection. 

When the stack spacings between the wide surfaces of the superconducting strips are large ($b/a>2$), we have $k_{\rm r}\simeq 4\exp(-\pi b/2a)\ll 1$, $\kappa_{\rm r}\simeq k_{\rm r}\cos(\pi w/2a)\ll 1$, and $c_{\rm r}\simeq 2a/\pi$, and Eq.~\eqref{eq:<mu_y>} reduces to 
\begin{equation}
	\frac{\mu_{\rm\perp r}}{\mu_0} 
	\simeq \left[ 1-\frac{2a}{\pi b}%
		\ln\cos\left(\frac{\pi w}{2a}\right)\right]^{-1} . 
\label{eq:<mu_y>large-stack-spacing}
\end{equation}

When the stack spacings between the wide surfaces of the superconducting strips are small ($b/a\ll 1$), we have $k_{\rm r}\simeq 1-8\exp(-\pi a/2b)$ and $\kappa_{\rm r}\simeq 1-2\exp[-(\pi a/b)(1-w/a)]$, and Eq.~\eqref{eq:<mu_y>} reduces to 
\begin{equation}
	\frac{\mu_{\rm\perp r}}{\mu_0}
	\simeq 1-\frac{w}{a}+\frac{2 b}{\pi a}\ln 2 , 
\label{eq:<mu_y>_small-stack-spacing}
\end{equation}
which is valid for a wide range of $w/a$. 
The linear behavior, $\mu_{\rm \perp r}/\mu_0 \to 1-w/a$, occurs because the rectangular array of strips with $b/a\to 0$ corresponds to superconducting slabs of width $2w$.~\cite{mawatari96prb54} 
Note that Eq.~\eqref{eq:<mu_y>_small-stack-spacing} is invalid for $w/a\ll 1$ or $1-w/a\ll 1$. 
When $b/a\ll 1$ and $1-w/a\ll 1$, Eq.~\eqref{eq:<mu_y>_small-gap} further reduces to 
\begin{equation}
	\frac{\mu_{\rm\perp r}}{\mu_0}
	\simeq \frac{\pi b}{2 a}\left[\ln\left(%
		\frac{8b}{\pi(a-w)} \right)\right]^{-1} . 
\label{eq:<mu_y>_small-gap-spacing}
\end{equation}

\begin{figure}[bt]
	\includegraphics{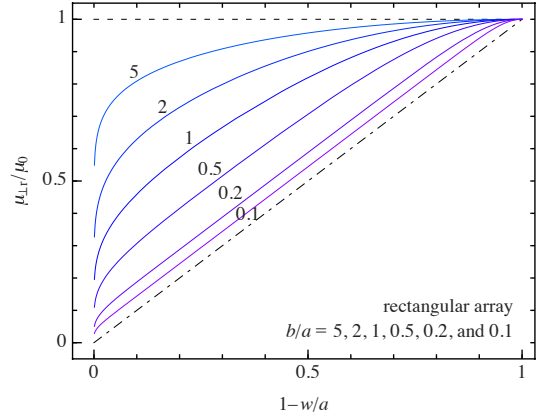}
\caption{(Color online) %
Effective permeability for perpendicular field $\mu_{\rm\perp r}$ as function of $1-w/a$ for $b/a= 5,\, 2,\, 1,\, 0.5,\, 0.2$, and $0.1$. 
The dashed line is $\mu_{\rm\perp r}/\mu_0=1$ for $b/a\to\infty$, and the dot-dash line is $\mu_{\rm\perp r}/\mu_0=1-w/a$ for $b/a\to 0$. 
}
\label{fig:mu-perp}
\end{figure}

Figure~\ref{fig:mu-perp} shows a plot of $\mu_{\rm\perp r}/\mu_0$ vs $1-w/a$ obtained from Eqs.~\eqref{eq:b/a-K}, \eqref{eq:k2}, and \eqref{eq:<mu_y>}. 

\begin{figure*}[bt]
	\includegraphics{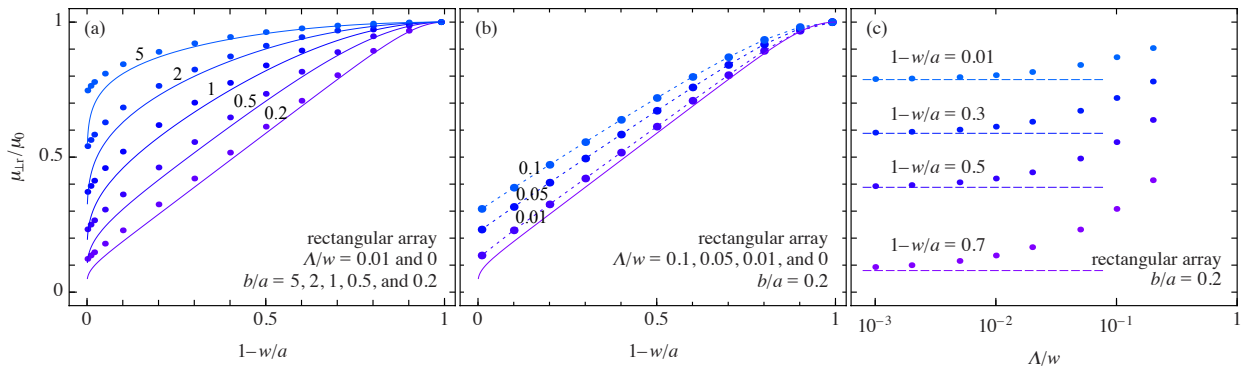}
\caption{(Color online) %
Comparison between the numerical results of the effective permeability $\mu_{\rm\perp r}$ in a rectangular array for $\Lambda/w>0$ (symbols) and the analytical results for $\Lambda/w\to 0$ (lines). 
(a) Numerical results (dots) for $\Lambda/w=0.01$ and analytical results (lines) for $\Lambda/w\to 0$ of $\mu_{\rm\perp r}/\mu_0$ vs $1-w/a$ for $b/a=5,\,2,\,1,\,0.5$, and $0.2$. 
(b) Numerical results (symbols) for $\Lambda/w=0.1,\,0.05,\,0.01$ and analytical results (line) for $\Lambda/w\to 0$ of $\mu_{\rm\perp r}/\mu_0$ vs $1-w/a$ for $b/a=0.2$. 
(c) Numerical results (dots) and analytical results (dashed lines) for $\Lambda/w\to 0$ of $\mu_{\rm\perp r}/\mu_0$ vs $\Lambda/w$ for $b/a=0.2$. 
}
\label{fig:numerical-mu_rect}
\end{figure*}

\subsection{Finite penetration depth, $\Lambda/w>0$}
We have performed numerical calculations for the realistic case taking nonzero $\Lambda/w$ into account. 
Our calculation of the sheet current in superconducting strips is based on the magnetic energy minimization technique, as described in Ref.~\onlinecite{navau09apl94}.

Figure~\ref{fig:numerical-mu_rect} shows a comparison between the numerical results of $\mu_{\rm\perp r}$ for $\Lambda/w>0$ and the analytical results for $\Lambda/w\to 0$. 
When $\Lambda/w\to 0$, the numerical results coincide with the analytical ones within the numerical accuracy. 
As shown in Figs.~\ref{fig:numerical-mu_rect}(a) and \ref{fig:numerical-mu_rect}(b), $\mu_{\rm\perp r}$ increases with increasing $\Lambda/w$ for any given values of $b/a$ and $1-w/a$, 
similar to what was found in Ref.~\onlinecite{navau09apl94}. 
When the gaps between the edges of the superconducting strips are small ($1-w/a<0.1$), we see a pronounced difference between $\mu_{\rm\perp r}/\mu_0$ for $\Lambda/w>0$ and that for $\Lambda/w\to 0$. 
We see a sharp drop in $\mu_{\rm\perp r}/\mu_0\to 0$ as $1-w/a\to 0$ for $\Lambda/w\to 0$ [see analytical expressions, Eqs.~\eqref{eq:<mu_y>_small-gap} and \eqref{eq:<mu_y>_small-gap-spacing}], in contrast to the slow decrease in $\mu_{\rm\perp r}/\mu_0$ with decreasing $1-w/a$ for $\Lambda/w=0.01$, because of the finite penetration of the magnetic field from the strip edges. 
Figure~\ref{fig:numerical-mu_rect}(c) shows $\mu_{\rm\perp r}$ as functions of $\Lambda/w$: the $\mu_{\rm\perp r}$ for $\Lambda/w<10^{-3}$ agree well with those for $\Lambda/w\to 0$, shown as dashed lines. 
We clearly see that $\mu_{\rm\perp r}$ for $\Lambda/w>0.1$ is much larger than that for $\Lambda/w\to 0$ especially for small $1-w/a$.

\section{Hexagonal array of superconducting strips
\label{sec:hexagonal-array}}

\begin{figure}[bt]
	\includegraphics{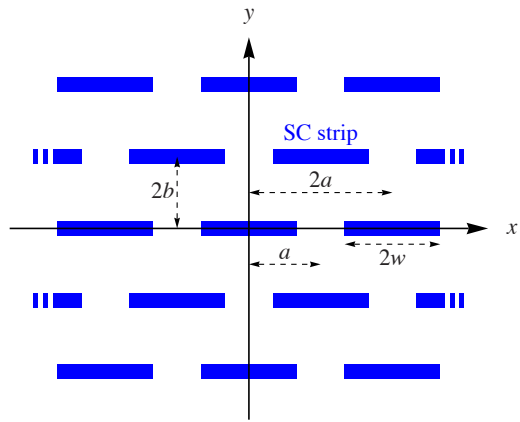}
\caption{(Color online) %
Cross section of a hexagonal array of superconducting strips in the $xy$ plane. 
In the even layer at $y=4nb$, the $m$th strip is at $|x-2ma|<w$, whereas in the odd layer at $y=(4n+2)b$, the $m$th strip is at $|x-(2m+1)a|<w$, where $m=0,\,\pm 1,\,\pm 2,\ldots\pm\infty$ and $n=0,\,\pm 1,\,\pm 2\ldots\pm\infty$. 
}
\label{fig:2D-hexag-array}
\end{figure}

In this section, we consider a hexagonal array of superconducting strips shown in Fig.~\ref{fig:2D-hexag-array}, and compare its response with that of the rectangular array shown in Fig.~\ref{fig:rectangular-array}. 

\subsection{Complete shielding state, $\Lambda/w\to 0$}
In this subsection we consider field distributions and perpendicular permeability of a hexagonal array of superconducting strips for $\Lambda/w\to 0$. 

\begin{figure*}[bt]
	\includegraphics{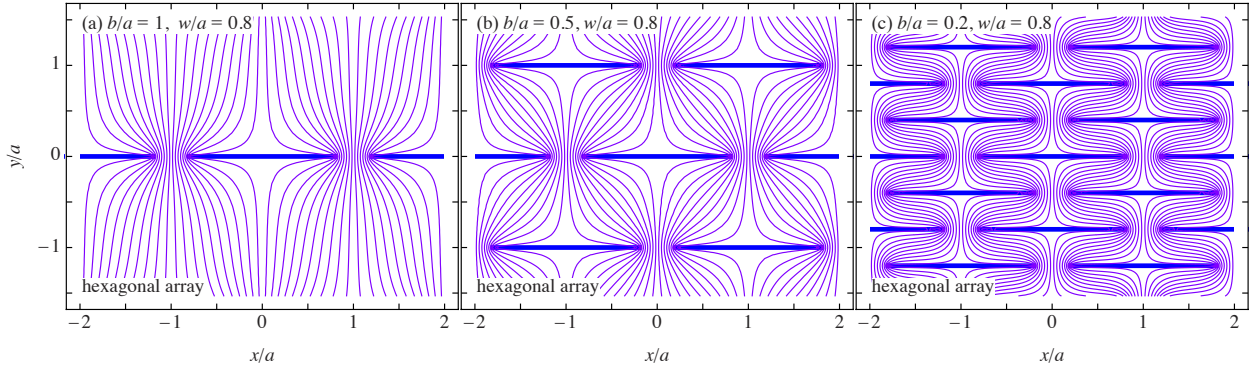}
\caption{(Color online) %
Magnetic field lines in a hexagonal array of superconducting strips for $w/a=0.8$: (a) $b/a=1$, (b) $b/a=0.5$, and (c) $b/a=0.2$. 
Thick horizontal lines correspond to the cross sections of superconducting strips. 
}
\label{fig:field-lines_hexag}
\end{figure*}

\subsubsection{Field distribution}
Here we use the conformal mapping from the $\zeta=x+iy$ plane to the $\eta=u+iv$ plane:~\cite{Kober} 
\begin{equation}
	\eta=\eta_{\rm h}(\zeta)\equiv \mbox{sn}(\zeta/c_{\rm h},k_{\rm h}) . 
\label{eq:zeta-eta_hexag}
\end{equation}
The modulus $k_{\rm h}$ is obtained by solving 
\begin{equation}
	\frac{2b}{a}= \frac{{\bf K}(k_{\rm h}')}{{\bf K}(k_{\rm h})} , 
\label{eq:2b/a-K3}
\end{equation}
where $k_{\rm h}'=\sqrt{1-k_{\rm h}^2}$. 
The relation between $k_{\rm h}$ from Eq.~\eqref{eq:2b/a-K3} and $k_{\rm r}$ from Eq.~\eqref{eq:b/a-K} is given by $k_{\rm h}=(1-k_{\rm r}')/(1+k_{\rm r}')=(1-\sqrt{1-k_{\rm r}^2})/(1+\sqrt{1-k_{\rm r}^2})$. 
$c_{\rm h}$ in Eq.~\eqref{eq:zeta-eta_hexag} is given by 
\begin{equation}
	c_{\rm h} =a/{\bf K}(k_{\rm h}) =2b/{\bf K}(k_{\rm h}') . 
\label{eq:c=a/K=b/K'_hexag}
\end{equation}
The upper half of the unit cell in the $\zeta$ plane, $|x|<a$ and $0<y<2b$, is mapped onto the upper half plane in the $\eta$ plane, $0<v<\infty$.~\cite{Kober} 
The points $\zeta=0$, $a$, $a+2ib$, and $2ib$ in the $\zeta$ plane are mapped onto $\eta=0$, $1$, $1/k_{\rm h}$, and $\infty$ in the $\eta$ plane, respectively.

The complex field ${\cal H}(\zeta)=H_y(x,y) +iH_x(x,y)$ describing the local field distribution for a hexagonal array in the complete shielding state is given by 
\begin{equation}
	{\cal H}(\zeta)= H_0\frac{\eta_{\rm h}(\zeta) 
		\sqrt{\eta_{\rm h}(\zeta)^2-k_{\rm h}^{-2}}}{%
		\sqrt{\left[\eta_{\rm h}(\zeta)^2- \gamma_{\rm h}^2\right] 
		\left[\eta_{\rm h}(\zeta)^2- \beta_{\rm h}^2\right]}} , 
\label{eq:H(z)_hexag}
\end{equation}
where $\gamma_{\rm h}=\eta_{\rm h}(w)=\mbox{sn}(w/c_{\rm h},k_{\rm h})$, and $\beta_{\rm h}=\eta_{\rm h}(a-w+2ib)=\sqrt{(k_{\rm h}^{-2}-\gamma_{\rm h}^2)/(1-\gamma_{\rm h}^2)}=(1/k_{\rm h})\mbox{dc}(w/c_{\rm h},k_{\rm h})$. 
(See Appendix A for Jacobi dc function.) 
Note that the coefficient $H_0=H_y(0,2b)$ in Eq.~\eqref{eq:H(z)_hexag} is the magnetic field at $(x,y)=(0,2b)$. 
For $b/a\to 0$, we have $k_{\rm h}\to 1$ and $\eta_{\rm h}(\zeta)\to\tanh(\pi\zeta/4b)$, which corresponds to the case of an infinite stack of superconducting strips.~\cite{mawatari96prb54} 
For $b/a\to \infty$, we have $k_{\rm h}\to 0$ and $\eta_{\rm h}(\zeta)\to\sin(\pi\zeta/2a)$, which corresponds to the case of an infinite array of coplanar superconducting strips.~\cite{mawatari96prb54} 

The perpendicular field component $H_y(x,0)={\rm Re}\,{\cal H}(x)$ and the sheet current $K_z(x)=H_x(x,-\epsilon)-H_x(x,\epsilon)={\rm Im}\,[{\cal H}(x-i\epsilon)-{\cal H}(x+i\epsilon)]$ 
are obtained from Eq.~\eqref{eq:H(z)_hexag}, and are given by Eqs.~\eqref{eq:hy(x,0)} and \eqref{eq:Kz(x)} by replacing $\eta_{\rm r}(x)$ by $\eta_{\rm h}(x)=\mbox{sn}(w/c_{\rm h},k_{\rm h})$. 

The complex potential defined by ${\cal G}(\zeta)= \int_{2ib}^{\zeta}{\cal H}(\zeta')d\zeta'$ is calculated from Eq.~\eqref{eq:H(z)_hexag}:
\begin{eqnarray}
	{\cal G}(\zeta) &=&
	\frac{c_{\rm h}H_0}{k_{\rm h}} \int_{\eta_{\rm h}(\zeta)}^{\infty} 
		\frac{udu}{\sqrt{(u^2-1)(u^2-\beta_{\rm h}^2)(u^2-\gamma_{\rm h}^2)}} 
\nonumber\\
	&=& \frac{c_{\rm h}H_0}{k_{\rm h}\sqrt{\beta_{\rm h}^2-\gamma_{\rm h}^2}} 
		F\left(\arcsin\sqrt{\frac{\beta_{\rm h}^2 
		-\gamma_{\rm h}^2}{\eta_{\rm h}(\zeta)^2-\gamma_{\rm h}^2}}, 
		\kappa_{\rm h}\right) , 
	\quad
\label{eq:G(z)_hexag}
\end{eqnarray}
where $\kappa_{\rm h}'=\sqrt{1-\kappa_{\rm h}^2}$ and $\kappa_{\rm h}$ is defined by 
\begin{equation}
	\kappa_{\rm h}= \sqrt{\frac{(1-\gamma_{\rm h}^2)^2}{%
		k_{\rm h}^{-2}-1+(1-\gamma_{\rm h}^2)^2}} 
	= \left[ 1+\frac{k_{\rm h}^{-2}-1}{%
		\mbox{cn}^4\,(w/c_{\rm h},k_{\rm h})} \right]^{-1/2} . 
\label{eq:k4}
\end{equation}
Figure~\ref{fig:field-lines_hexag} shows the magnetic field lines that correspond to the contour lines of ${\rm Re}\,{\cal G}(x+iy)$ obtained from Eq.~\eqref{eq:G(z)_hexag}. 
When the stack spacing is large ($b/a\gtrsim 1$), field distributions in a hexagonal array are similar to those in a rectangular array, as seen in Figs.~\ref{fig:field-lines}(a) and \ref{fig:field-lines_hexag}(a). 
When the stack spacing is small ($b/a<1$), on the other hand, we see a striking difference between the rectangular and hexagonal arrays. 
The magnetic field concentrates only near the gaps between the edges of the strips in a rectangular array [Fig.~\ref{fig:field-lines}(c)], whereas the magnetic field is large in most of the region in a hexagonal array [Fig.~\ref{fig:field-lines_hexag}(c)].

\subsubsection{Macroscopic fields}
The definitions of the macroscopic magnetic induction $\langle B_y\rangle$ and the magnetization $\langle M_y\rangle$ for a hexagonal array are the same as those of a rectangular array, and are given by Eqs.~\eqref{eq:<By>_def} and \eqref{eq:My}, respectively. 

The definition of the macroscopic magnetic field $\langle H_y\rangle$ given by Eq.~\eqref{eq:<Hy>_def} for a rectangular array, on the other hand, must be modified so that $\langle H_y\rangle$ for the hexagonal array is consistent with the macroscopic relation given by Eq.~\eqref{eq:By-Hy=My}. 
We therefore define $\langle H_y\rangle$ for a hexagonal array as 
\begin{equation}
	\langle H_y\rangle \equiv 
	\frac{1}{2b} \left[ \int_0^{2b} H_y(a,y)dy 
	-\int_0^a H_x(x,2b-\epsilon)dx \right] . 
\label{eq:hexag_<Hy>_def}
\end{equation}

\begin{figure}[bt]
	\includegraphics{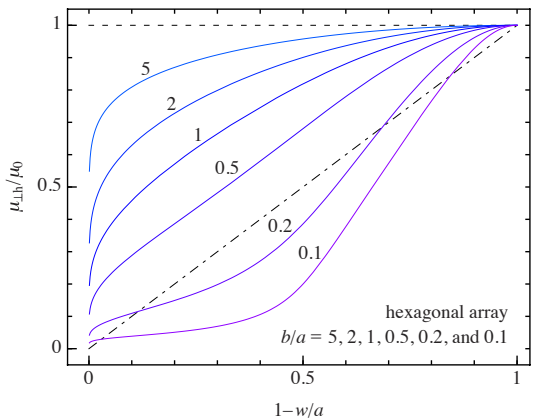}
\caption{(Color online) %
Effective permeability for perpendicular field $\mu_{\rm\perp h}$ as function of $1-w/a$ for $b/a= 5,\, 2,\, 1,\, 0.5,\, 0.2$, and $0.1$. 
The dashed line is $\mu_{\rm\perp h}/\mu_0=1$ for $b/a\to\infty$. The dot-dash line of $\mu_{\perp}/\mu_0=1-w/a$ is shown for comparison with Fig.~\ref{fig:mu-perp}. 
}
\label{fig:mu-perp_hexag}
\end{figure}

\subsubsection{Perpendicular permeability}
The resulting effective permeability for hexagonal array $\mu_{\rm\perp h}=\langle B_y \rangle/{\langle H_y \rangle}$ is given by 
\begin{equation}
	\frac{\mu_{\rm\perp h}}{\mu_0} 
	= \frac{2b}{a} \frac{{\bf K}(\kappa_{\rm h})}{%
		{\bf K}(\kappa_{\rm h}')} , 
\label{eq:<mu_y>_hexag}
\end{equation}
where $\kappa_{\rm h}$ is defined by Eq.~\eqref{eq:k4}. 

When the width of superconducting strips is small ($w/c_{\rm h}\ll 1$), we have $\kappa_{\rm h}\simeq k_{\rm h}[1-(1-k_{\rm h}^2)w^2/2c_{\rm h}^2]$, and Eq.~\eqref{eq:<mu_y>_hexag} reduces to Eq.~\eqref{eq:<mu_y>_small-width}. 

When the gaps between the edges of the superconducting strips are small ($1-w/a\ll 1$), we have 
\begin{equation}
	\frac{\mu_{\rm\perp h}}{\mu_0}
	\simeq \frac{\pi b}{2a}\left[\ln\left(%
		\frac{2}{[k_{\rm h}^2(1-k_{\rm h}^2)]^{1/4}{\bf K}(k_{\rm h})(1-w/a)} \right)\right]^{-1} . 
\label{eq:<mu_y>_small-gap_hexag}
\end{equation}
Equation~\eqref{eq:<mu_y>_small-gap_hexag} is further simplified for $b/a\ll 1$ as 
\begin{equation}
	\frac{\mu_{\rm\perp h}}{\mu_0}
	\simeq \frac{\pi b}{2a}\left[ \frac{\pi a}{8 b} 
		+\ln\left( \frac{4 b}{\pi(a-w)} \right)\right]^{-1} . 
\label{eq:<mu_y>_small-gap-spacing_hexag}
\end{equation}

When the stack spacings between the wide surfaces of the superconducting strips are large ($b/a>2$), we have $k_{\rm h}\simeq 4\exp(-\pi b/a)\ll 1$, $\kappa_{\rm h}\simeq k_{\rm h}\cos^2(\pi w/2a)\ll 1$, and $c_{\rm h}\simeq 2a/\pi$, and Eq.~\eqref{eq:<mu_y>_hexag} reduces to Eq.~\eqref{eq:<mu_y>large-stack-spacing}. 
In other words, the magnetic response of the hexagonal array is almost the same as that of the rectangular array for large $b/a$, as expected.

Figure~\ref{fig:mu-perp_hexag} shows a plot of $\mu_{\rm \perp h}/\mu_0$ vs $1-w/a$ obtained from Eqs.~\eqref{eq:2b/a-K3}, \eqref{eq:k4}, and \eqref{eq:<mu_y>_hexag}. 
Note that for a given $b/a$, $\mu_{\rm\perp h}$ of the hexagonal array shown in Fig.~\ref{fig:2D-hexag-array} is smaller than $\mu_{\rm\perp h}$ of the rectangular array shown in Fig.~\ref{fig:rectangular-array}, especially when $b/a\ll 1$. 
The effective permeability of the hexagonal array is very small and slightly dependent on the gaps between the edges of the superconducting strips; $\mu_{\rm\perp h}/\mu_0\ll 1$ when the stack spacing is small ($b/a\ll 1$) for the wide range of $w/a$, $0<1-w/a\lesssim 0.5$. 
The hexagonal array is, therefore, more advantageous in obtaining highly anisotropic permeability than the rectangular array. 

\begin{figure*}[bt]
	\includegraphics{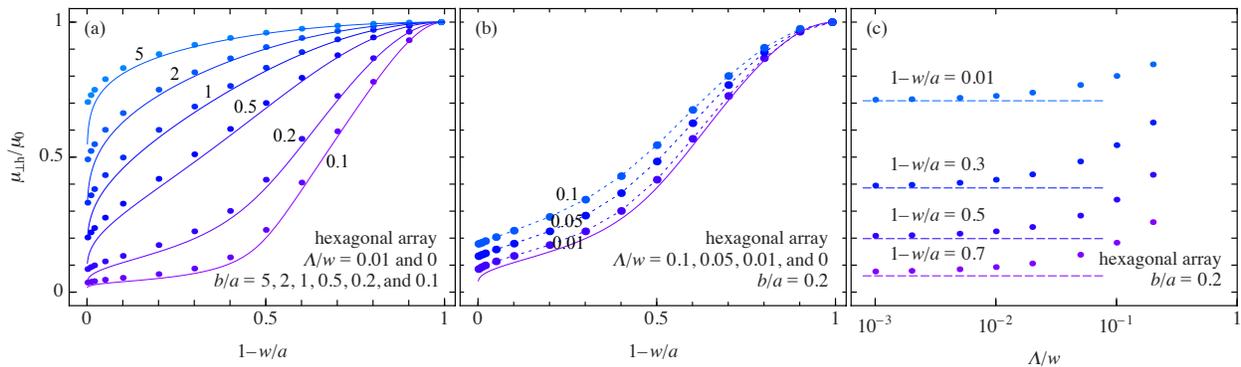}
\caption{(Color online) %
Comparison between the numerical results of the effective permeability $\mu_{\rm\perp h}$ in a hexagonal array for $\Lambda/w>0$ (symbols) and the analytical results for $\Lambda/w\to 0$ (lines). 
(a) Numerical results (dots) for $\Lambda/w=0.01$ and analytical results (lines) for $\Lambda/w\to 0$ of $\mu_{\rm\perp h}/\mu_0$ vs $1-w/a$ for $b/a=5,\,2,\,1,\,0.5,\,0.2$, and $0.1$. 
(b) Numerical results (symbols) for $\Lambda/w=0.1,\,0.05,\,0.01$ and analytical results (line) for $\Lambda/w\to 0$ of $\mu_{\rm\perp h}/\mu_0$ vs $1-w/a$ for $b/a=0.2$. 
(c) Numerical results (dots) and analytical results (dashed lines) for $\Lambda/w\to 0$ of $\mu_{\rm\perp h}/\mu_0$ vs $\Lambda/w$ for $b/a=0.2$. 
}
\label{fig:numerical-mu_hexag}
\end{figure*}

\subsection{Finite penetration depth, $\Lambda/w>0$}
Figure~\ref{fig:numerical-mu_hexag} shows a comparison between the numerical results of $\mu_{\rm\perp h}$ for $\Lambda/w>0$ and the analytical results for $\Lambda/w\to 0$. 
The overall tendency in the difference between $\mu_{\rm\perp h}$ for $\Lambda/w>0$ and that for $\Lambda/w\to 0$ shown in Fig.~\ref{fig:numerical-mu_hexag} is similar to that in Fig.~\ref{fig:numerical-mu_rect}. 
It is interesting to compare $\mu_{\rm\perp h}$ of a hexagonal array and $\mu_{\rm\perp r}$ of a rectangular array for small stack spacings ($b/a\ll 1$), in that the $\mu_{\rm\perp h}$ for $\Lambda/w>0$ is close to that for $\Lambda/w\to 0$, even for small $1-w/a$.

\section{Discussion and Summary
\label{sec:discussion}}
In this paper, we have investigated anisotropic effective permeability of infinite arrays of thin superconducting strips, which can be regarded as essential structures of a dc magnetic metamaterial. 
When an applied magnetic field is parallel to the superconducting strips, the effective permeability is given by $\mu_{\parallel}=\mu_0$. 
When an applied magnetic field is perpendicular to the strip, on the other hand, the effective permeability is $\mu_{\perp}<\mu_0$, because of the magnetic shielding in the superconducting strips. 
The perpendicular permeability $\mu_{\perp}$ becomes small when the stack spacings are small ($b/a\ll 1$), or when the gaps between the edges of the strips are small ($1-w/a\ll 1$). 
We have investigated two types of 2D arrays: the rectangular array shown in Fig.~\ref{fig:rectangular-array} and the hexagonal array shown in Fig.~\ref{fig:2D-hexag-array}. 
The hexagonal array is more advantageous for obtaining large anisotropy of the effective permeability than the rectangular array, because $\mu_{\rm\perp h}$ is much smaller than $\mu_{\rm\perp r}$ when $b/a\ll 1$ and $1-w/a\lesssim 0.5$. 

To realize magnetic cloaking or other possible applications for controlling magnetic fields, magnetic metamaterials with large anisotropy ($\mu_{\perp}/\mu_0\ll 1$) are needed. 
As seen in Figs.~\ref{fig:mu-perp} and \ref{fig:mu-perp_hexag}, $\mu_{\perp}/\mu_0\ll 1$ can be achieved by means of an array of superconducting strips with small gaps $1-w/a\ll 1$ and small stack spacing $b/a\ll 1$. 
For finite $\Lambda/w$, however, $\mu_{\perp}/\mu_0\ll 1$ is difficult to realize, because of field penetration near the edges of the superconducting strips. 
Wide superconducting strips are favorable for obtaining small $\Lambda/w$ and small $\mu_{\perp}/\mu_0$, but control of magnetic fields for magnetic cloaking, for example, requires narrow superconducting strips. 
We therefore need to consider optimization of the strip size and configuration of the array of superconducting strips. Such optimization depends on the details of the application for magnetic metamaterials.

\section*{ACKNOWLEDGMENTS} 
We thank D.-X.\ Chen and N.\ Del-Valle for helpful discussions. 
C.\ N.\ and A.\ S.\ acknowledge Consolider Project NANOSELECT (CSD2007-00041) for financial support.

\appendix 
\section{Elliptic integrals and functions
\label{sec:elliptic-function}}
In this paper, some elliptic integrals and functions appear, but those notations and definitions are not unified in the literature. 
In this appendix, we therefore summarize those expressions to avoid ambiguity. The expressions below are from the textbook by Gradsytein and Ryshik.~\cite{Gradshtein-Ryzhik} 

The elliptic integral of the first kind $F(\varphi,k)$ and the complete elliptic integral of the first kind ${\bf K}(k)$ are given by 
\begin{eqnarray}
	F(\varphi,k) &=& \int_0^{\sin\varphi}
		\frac{dt}{\sqrt{(1-t^2)(1-k^2t^2)}} , 
\label{eq:F_elliptic-int}\\
	{\bf K}(k) &=& F(\pi/2,k) 
	= \int_0^1 \frac{dt}{\sqrt{(1-t^2)(1-k^2t^2)}} , 
	\qquad
\label{eq:K_comp-ellip-int}
\end{eqnarray}
respectively. 

The Jacobian elliptic functions $\mbox{sn}(u,k)$ (i.e., sine amplitude), $\mbox{cn}(u,k)$ (i.e., cosine amplitude), and $\mbox{dn}(u,k)$ (i.e., delta amplitude) are defined via $u=F(\varphi,k)$ by 
\begin{eqnarray}
	\mathrm{sn}(u,k) &=& \sin\varphi , 
\label{eq:sn}\\
	\mathrm{cn}(u,k) &=& \cos\varphi , 
\label{eq:cn}\\
	\mathrm{dn}(u,k) &=& \sqrt{1-k^2\sin^2\varphi} , 
\label{eq:dn}
\end{eqnarray}
respectively. 
We also use $\mbox{cd}(u,k)=\mbox{cn}(u,k)/\mbox{dn}(u,k)$ and $\mbox{dc}(u,k)=\mbox{dn}(u,k)/\mbox{cn}(u,k)=1/\mbox{cd}(u,k)$. 

Here are simple expressions of $\mbox{sn}(u,k)$ for specific $k$ or $u$: 
\begin{eqnarray}
	\mbox{sn}(u,1) &=& \tanh(u) , 
\label{eq:sn_k=1}\\
	\mbox{sn}(u,0) &=& \sin(u) . 
\label{eq:sn_k=0}\\
	\mbox{sn}(0,k) &=& 0 , 
\label{eq:sn_u=0}\\
	\mbox{sn}({\bf K},k) &=& 1 , 
\label{eq:sn_u=K}\\
	\mbox{sn}({\bf K}+i{\bf K}',k) &=& 1/k , 
\label{eq:sn_u=K+iK'}\\
	\mbox{sn}(i{\bf K}',k) &=& \infty , 
\label{eq:sn_u=iK'}
\end{eqnarray}
where ${\bf K}={\bf K}(k)$ and ${\bf K}'={\bf K}(k')={\bf K}(\sqrt{1-k^2})$.

\end{document}